\documentclass[11pt,a4paper]{article}   
\usepackage{amsmath}
\usepackage{amssymb}
\usepackage{mathrsfs}

\usepackage{rotating}

\usepackage[round]{natbib}

\newcommand{\eg}{e.g.\ }
\newcommand{\ie}{i.e.\ }

\newcommand{\cvar}{{\rm i}}
\newcommand{\expo}{{\rm e}}
\newcommand{\ud}{~{\rm d}}
\newcommand{\besselj}{{\rm J}}
\newcommand{\besselh}{{\rm H}}

\begin{document}

\title{Reflection and transmission of ocean wave spectra by a band of randomly distributed ice floes} 

\author{Fabien MONTIEL,$^1$
  Vernon A. SQUIRE,$^1$ Luke G. BENNETTS$^2$
\\
{\footnotesize
$^1$Department of Mathematics and Statistics,
  University of Otago,
  PO Box 56, Dunedin 9054, New Zealand}
\\
{\footnotesize
$^2$School of Mathematical Sciences,
  University of Adelaide,
  Adelaide, South Australia 5005, Australia}}
  
\date{\today} 
\maketitle 

\abstract{
A new ocean wave/sea-ice interaction model is proposed that simulates how a directional wave spectrum evolves as it travels through an arbitrary finite 
array of circular ice floes, where wave/ice dynamics are entirely governed by wave scattering effects. The model is applied to characterise the wave 
reflection and transmission properties of a strip of ice floes, such as an ice edge band. A method is devised to extract the reflected and transmitted 
directional wave spectra produced by the array. The method builds upon an integral mapping from polar to Cartesian coordinates of the scattered wave 
components. Sensitivity tests are conducted for a row of floes randomly perturbed from a regular arrangement. Results for random arrays are generated using 
ensemble averaging. A realistic ice edge band is then reconstructed from field experiments data. Simulations show a good qualitative agreement with the 
data in terms of transmitted wave energy and directional spreading. In particular, it is observed that short waves become isotropic quickly after 
penetrating the ice field. 
}

\section{Introduction}

The polar oceans are changing in response to climate change, particularly the Arctic sea ice cover during the summer season. Over the last few decades, 
Arctic sea ice has thinned, shifted from predominately perennial ice to seasonal (first-year) ice, and declined in extent at a rate of $-13\%$ per decade 
relative to the 1979--2000 average at the end of the summer melt period \citep{jeffries_etal13}. The resulting increase in the absorption of short wave 
radiation by the ocean tends to induce additional melt and forms the basis of the positive ice albedo-temperature feedback. 
Severe storms, such as that of early August 2012, which destroyed 400,000 square km of sea ice in just a few days \citep{parkinson_comiso13}, are also 
occurring more often at high latitudes in both hemispheres, with increased fetches apparently triggering amplified wave activity within the marginal seas 
and along newly exposed coastal borders \citep{young_etal11}. It is therefore crucial to determine if a changing wave regime could further enhance ice 
retreat and thinning through increased melt rates due to floe breakage and water movements, acting as an additional positive influence on ice 
albedo-temperature feedback.

Driven by these observations there has been a resurgence of interest in understanding how ocean waves interact with sea ice. The goal is to improve the 
accuracy of ice/ocean models, which have been unable to predict the recent fast decline of the Arctic sea ice cover \citep{jeffries_etal13}. The so called 
WIFAR (Waves-in-Ice Forecasting for Arctic Operators) project reported briefly by \cite{squire_etal13} and in more detail by 
\cite{williams_etal13a,williams_etal13b} is a first attempt to advect ocean waves into a coupled ice/ocean model. Momentum exchanges between ocean waves 
and sea ice are largely based on conservative physics, where ocean wave scattering by ice floes governs the dynamics of the system. This is achieved using 
the methods described by \cite{bennetts_squire12}, who consider wave propagation in a single horizontal dimension. While several wave vectors at different 
angles can be accommodated, they are independent so that transmission and reflection by ice floes is constrained to a single direction. This is a major 
simplification that is reasonable deep into the ice interior \citep{squire_etal09}, but is likely to be imprecise within the marginal ice zone (MIZ) 
where the waves are thought to be multiply scattered in all directions by the constituent ice floes \citep{wadhams_etal86}. Each floe present will 
produce circular wave fronts that interact with the floes around it, to quickly produce a confused sea state that tends towards being isotropic in its 
directional composition --- particularly when wave periods are low. Although models of multiple 2D scattering have been developed in the past 
\citep{bennetts_squire09, bennetts_etal10}, these have had unrealistic restraints imposed, e.g.\ periodicity, so a more general theoretical framework is 
needed.

In this paper we describe such a framework, with no periodicity limitation and full randomisation of the floe size distribution (FSD). We use it to 
characterise the wave reflection and transmission properties by a row composed of O(10) circular floes, after averaging over many random realizations of 
the FSD, allowing us to eliminate features that relate to a particular realization. Our approach extends the single row models proposed by 
\cite{peter_etal06} and \cite{bennetts_squire08} which are based on a periodicity condition that causes the scattered waves to only travel at a finite 
number of angles determined by the floe spacing. For an arbitrary array, however, the full directional spectrum must be considered. The main 
mathematical challenge is to compute reflection and transmission wave characteristics for scattering by circular floes. We use an integral transformation 
to recast the circular wave forms into plane waves travelling at different angles, utilising a method developed for electromagnetic wave scattering 
\citep{cincotti_etal93,frezza_etal10}. The approach is new for water waves travelling through an ice field and is particularly efficacious, 
as it allows the attenuation and directional evolution to be found simultaneously by dividing up the propagation medium into strips. The theoretical 
aspects of the method will be described in the following section.

Our model is especially suitable to replicate the behaviour of ice edge bands, i.e.\ consolidated structures that form off the ice edge of the MIZ, 
which received considerable attention during MIZEX \citep{martin_etal83,wadhams83}. While the origin and causal mechanism of ice edge bands are reasonably 
well understood \citep{wadhams83}, few data exist that quantify band interactions with penetrating wave trains. Accordingly, this work is of geophysical 
interest, as bands are a frequent resident off the ice edge of certain MIZs. Useful data regarding the FSD of ice bands are provided by 
\cite{wadhams_etal86}, which we use to parametrise realistic ice conditions and simulate the propagation of a directional wave spectrum through the 
band. 

After describing the numerical model in the next section, we conduct an analysis for a single row of floes with random perturbations from the regular 
array arrangement. We test the effect of randomising the FSD and increasing the extent of the row on the transmitted wave energy. Results for a realistic 
ice band are then given, and qualitative agreement with the experimental data reported by \cite{wadhams_etal86} is found.


\section{Model}

\begin{figure}
  \centering{\includegraphics[width=0.65\textwidth]{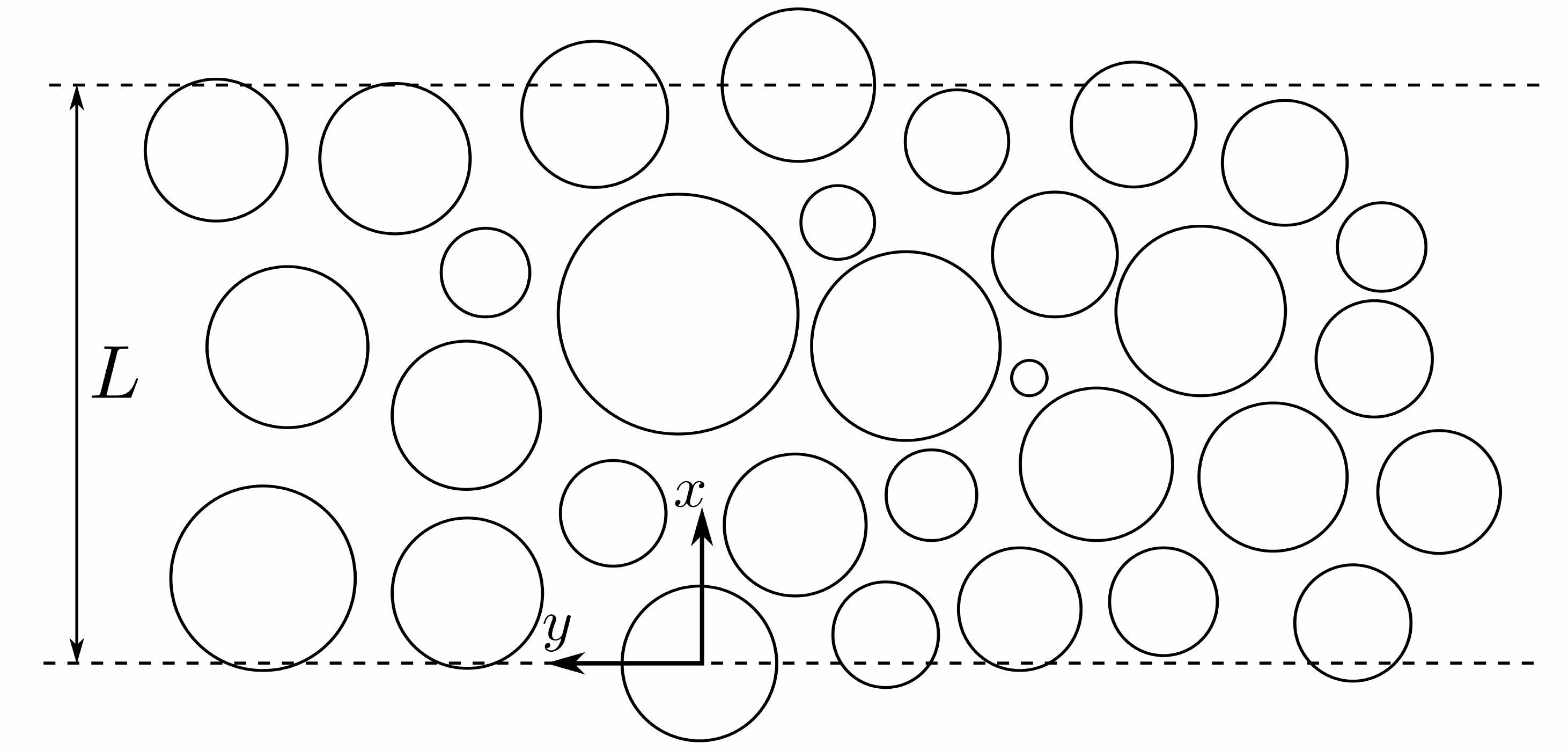}}
  \caption{Diagram of a typical band of ice floes with width $L$.}
  \label{fig1}
\end{figure}

Consider a coherent band of sea ice floes herded together at the surface of the ocean, which is assumed to have infinite horizontal extent and a finite 
constant depth $h$. Cartesian coordinates $\mathbf{x} = (x,y,z)$ are used to position points in the fluid domain, with the $z$-axis pointing vertically 
upwards from the origin located on the fluid surface at rest. We assume that the ice floes are circular and vertically uniform, although we let their 
size, i.e.\ radius and thickness, and their position be arbitrary, with the restriction that they do not overlap. (We conjecture that the idealised 
circular floe model represents a realistic ice cover if randomness is included and the results are averaged over many simulations.) It is supposed that 
wave/ice dynamics are governed entirely by conservative scattering effects, so that dissipative phenomena, \eg friction, floe collisions, rafting, ice 
fracture and overwash, are not considered at this stage. Figure~\ref{fig1} shows a diagram of a typical ice band FSD. The lowermost floe has its centre 
located at the origin, without loss of generality, and we define the width of the band, denoted by $L$, as the $x$-coordinate of the centre of the 
uppermost floe. We let $M$ denote the number of floes in the band and $(x_j,y_j)$ be the coordinates of floe $j$, $1\le j\le M$.

\subsection{Governing equations}

Under the standard assumptions of potential flow, for an incompressible fluid with density $\rho = 1025\rm\,kg\,m^{-3}$, and assuming periodic motion with 
angular frequency $\omega$, we describe the fluid motion using a potential function defined as the gradient of the velocity field. It is written as 
$\Phi(\mathbf{x},t) = \mathrm{Re}\{(g/\cvar\omega)\phi(\mathbf{x})\expo^{-\cvar \omega t}\}$, where $g\approx9.81\rm\,m\,s^{-2}$ is the acceleration due 
to gravity. The complex-valued function $\phi$ satisfies Laplace's equation 
\begin{equation}
  \left(\nabla^2 + \partial_z^2\right)\phi=0
\end{equation}
everywhere in the fluid, with $\nabla \equiv (\partial_x,\partial_y)$. 

Boundary conditions must be imposed on the seabed and on the fluid upper surface. Assuming no flow through the seabed, we have $\partial_z\phi=0$ on the 
plane $z=-h$. The fluid upper surface comprises both open water and ice-covered regions. We invoke linear water wave theory at the free surface, which 
provides the condition
\begin{equation}
  \partial_z\phi = \alpha\phi \qquad (z=0),
\end{equation}
valid in open water regions, where $\alpha=\omega^2/g$. We account for the flexural motion experienced by the ice floes under wave action, but neglect 
any form of horizontal motion. A standard hydroelastic model, based on thin-elastic plate theory, provides the corresponding condition
\begin{equation}
  \left(\beta\nabla^4 + 1-\alpha d\right)\partial_z\phi = \alpha\phi \qquad (z=-d),
\end{equation}
valid in the ice-covered fluid regions, for an ice floe with thickness $D$, density $\rho_i\approx922.5\rm\,kg\,m^{-3}$ and draught $d=(\rho_i/\rho)D$. The 
stiffness parameter $\beta=F/\rho g$ is defined in terms of the flexural rigidity $F = ED^3/12(1-\nu^2)$, where $E\approx6\rm\,GPa$ is the effective 
Young's modulus for sea-ice and $\nu\approx0.3$ Poisson's ratio. Further boundary conditions are applied that ensure that the bending moment and vertical 
shear stress of the floes vanish at their edges \citep[see, e.g.,][]{montiel_etal13b}. A radiation condition is also imposed in the far field to ensure the 
decay of scattered wave components. 

\subsection{Directional wave spectrum}

A key goal of the present study is to analyse how angular spreading is affected by a band of ice floes. We consider a monochromatic wave forcing with 
normalised angular distribution, incident on the ice band from the negative $x$-axis. We set the directional spread of the incident wave as a standard 
cosine-squared directional spectral density function $S(\tau)$, $-\pi/2\le\tau\le\pi/2$, that is
\begin{equation}
  S(\tau) = \frac{2}{\pi}\cos^2(\tau), \qquad \mathrm{such\,that} \qquad \int_{-\pi/2}^{\pi/2}S(\tau) \ud \tau = 1.
\end{equation}
The incident wave potential $\phi^{\mathrm{In}}$ can then be defined as a continuous superposition of plane waves travelling at different directions in 
$(-\pi/2,\pi/2)$. The following integral representation follows naturally as
\begin{equation}
  \phi^{\mathrm{In}}(\mathbf{x}) = \frac{\cosh k(z+h)}{\cosh kh} 
  \int_{-\pi/2}^{\pi/2} A^{\mathrm{In}}(\tau) \expo^{\cvar k (x\cos\tau + y\sin\tau)} \ud \tau,
  \label{eq5}
\end{equation}
where $k$ is the open water wavenumber, \ie\ the solution of the dispersion relation $k\tanh kh = \alpha$. The incident wave amplitude spectrum is defined 
as $A^{\mathrm{In}}(\tau) = \sqrt{S(\tau)}$. 

Given the representation of the wave field used in \eqref{eq5}, it is sensible to define the reflected and transmitted components that arise from 
scattering by the band, in the form
\begin{eqnarray}
  \phi^{\mathrm{R}}(x,y,z) & = & \frac{\cosh k(z+h)}{\cosh kh} \int_{-\pi/2}^{\pi/2} A^{\mathrm{R}}(\chi) 
  \expo^{\cvar k(-x\cos\chi + y\sin\chi)} \ud\chi \quad (x \le 0), \label{eq1} \label{eq6}\\ 
  \phi^{\mathrm{T}}(x,y,z) & = & \frac{\cosh k(z+h)}{\cosh kh} \int_{-\pi/2}^{\pi/2} A^{\mathrm{T}}(\chi) 
  \expo^{\cvar k((x-L)\cos\chi + y\sin\chi)} \ud\chi \quad (x \ge L),\label{eq2} \label{eq7}
\end{eqnarray}
where $A^{\mathrm{R}}(\chi)$ and $A^{\mathrm{T}}(\chi)$ are the reflection and transmission spectra, respectively, which are unknowns of the problem. We 
decompose further the transmitted spectrum into a scattered wave component $\widetilde{A}^{\mathrm{T}}(\chi)$ and the contribution from the ambient 
field, such that $A^{\mathrm{T}}(\chi) = \widetilde{A}^{\mathrm{T}}(\chi) + \expo^{\cvar kL\cos\chi}A^{\mathrm{In}}(\chi)$. Also note that we have 
neglected the effect of the evanescent wave components in these representations, as we are only interested in the reflected and transmitted energy.

\subsection{Solution method}

We first solve the multiple scattering problem in the band, using a standard interaction theory \citep[see, e.g.,][]{kagemoto_yue86}, which states that 
the wave field incident on a floe $j$, $1\le j\le M$, is the superposition of the ambient wave forcing and the scattered wave field from all the other 
floes present. We use local cylindrical coordinates $(r_j,\theta_j,z)$ to parametrise the incident and scattered wave fields for floe $j$. The ambient 
wave forcing given by equation \eqref{eq5} is then expressed in terms of cylindrical harmonic functions as
\begin{subequations}\label{eq8}
\begin{equation}
  \phi^{\mathrm{In}}(r_j,\theta_j,z) \approx \frac{\cosh k(z+h)}{\cosh kh} \sum_{n=-N}^{N} a_n^{(j)} \besselj_n(k r_j) \expo^{\cvar n\theta_j}, 
\end{equation}
\begin{equation}  
  \mathrm{with} \quad 
  a_n^{(j)}=\cvar^n\int_{-\pi/2}^{\pi/2}A^{\mathrm{In}}(\tau)\expo^{-\cvar n\tau}\expo^{\cvar k(x_j\cos\tau + y_j\sin\tau)} \ud\tau,
\end{equation}
\end{subequations}
where $\besselj_n$ is the Bessel function of the first kind of order $n$ and the infinite sum has been truncated to $2N+1$ terms for computational 
purposes. We also represent the scattered wave field for each floe $j$ in the form
\begin{equation}
  \phi^{\mathrm{S}}_j(r_j,\theta_j,z) \approx \frac{\cosh k(z+h)}{\cosh kh} \sum_{n=-N}^{N} b_n^{(j)} \besselh_n(k r_j) \expo^{\cvar n\theta_j},
  \label{eq9}
\end{equation}
assuming that the evanescent wave modes are negligible (wide-spacing approximation). 

At this point, solving the scattering problem for a single circular floe is necessary to map the total incident wave field on floe $j$ to the scattered 
wave field. We use a standard eigenfunction matching method, which provides reasonable numerical accuracy for the present problem. The method is 
described in \cite{montiel_etal13b}, and also provides a solution in the ice-covered region. Note that evanescent modes were included to compute the 
single floe responses with sufficient accuracy. 

Using the superposition principle from the interaction theory and Graf's addition theorem yields a mapping between the unknown amplitudes $b_n^{(j)}$ and 
the forcing amplitudes $a_n^{(j)}$, $-N\le n\le N$, $1\le j\le M$, given by the matrix equation \citep{montiel12}
\begin{equation}
  \mathbf{b} = \mathbf{\mathfrak{D}} \mathbf{a}.
  \label{eq10}
\end{equation}
The column vectors $\mathbf{a}$ and $\mathbf{b}$ have length $M(2N+1)$ and contain the amplitudes $a_n^{(j)}$ and $b_n^{(j)}$, respectively. The mapping 
matrix $\mathbf{\mathfrak{D}}$ is usually referred to as the diffraction transfer matrix of the array. 

The solution given by equations \eqref{eq9} and \eqref{eq10} is not compatible with the plane wave expression of the scattered field described by 
equations \eqref{eq6} and \eqref{eq7}. Therefore, we use a plane wave representation of the cylindrical harmonic functions 
$\besselh_n(kr_j)\exp(\cvar n\theta_j)$, to obtain the necessary change of coordinates. \cite{cincotti_etal93} gave such a representation in the context 
of electromagnetic wave scattering. Adapting their expression to the present context, we have
\begin{equation}
  \besselh_n(kr_j)\expo^{\cvar n\theta_j} = \left\{
  \begin{array}{ll}
    \displaystyle{\frac{\cvar^n}{\pi} \int_{-\pi/2+\cvar\infty}^{\pi/2-\cvar\infty} \expo^{-\cvar n\chi} 
    \expo^{\cvar k(-x\cos\chi + y\sin\chi)}\ud \chi,} & \quad (x \le x_j),\\
    \displaystyle{\frac{(-\cvar)^n}{\pi} \int_{-\pi/2+\cvar\infty}^{\pi/2-\cvar\infty} \expo^{\cvar n\chi} 
    \expo^{\cvar k(x\cos\chi + y\sin\chi)}\ud \chi,} & \quad (x \ge x_j),
  \end{array}
  \right.
  \label{eq11}
\end{equation}
for each floe $j$. These expressions are valid along infinite lines parallel to the $y$-axis (\ie\ for fixed $x$). In particular, we can express the 
reflected components from each floe on the line $x=0$ and the transmitted components on the line $x=L$. Also note that the range of integration in 
equation \eqref{eq11} spreads into the complex plane. It can be shown, however, that for complex $\chi$ the plane wave term decays exponentially. It is 
then reasonable to limit the integration range on the real line to $(-\pi/2,\pi/2)$, by neglecting the contribution of the evanescent components.

We are now in a position to express the scattered wave field in the form of equations \eqref{eq6} and \eqref{eq7}. Substituting \eqref{eq11} into 
\eqref{eq9} and using \eqref{eq10} and \eqref{eq8}, we can obtain the following integral mappings
\begin{equation}
  A^{\mathrm{R}}(\chi) = \int_{-\pi/2}^{\pi/2} \mathcal{R}(\chi:\tau) A^{\mathrm{In}}(\tau) \ud \tau \quad \mathrm{and} \quad
  \widetilde{A}^{\mathrm{T}}(\chi) = \int_{-\pi/2}^{\pi/2} \mathcal{T}(\chi:\tau) A^{\mathrm{In}}(\tau) \ud \tau,
  \label{eq12}
\end{equation}
where $\mathcal{R}(\chi:\tau)$ and $\mathcal{T}(\chi:\tau)$ are the reflection and transmission kernels, respectively. The kernel functions 
$\mathcal{R}$/$\mathcal{T}$ map an incident wave of angle $\tau$ to a reflected/transmitted wave travelling at angle $\chi$. After some algebra, we obtain 
analytical expression for the kernels
\begin{equation}
  \mathcal{R}(\chi:\tau) = \left(\mathbf{V}^{\mathrm{R}}(\chi)\right)^\mathrm{tr} \mathbf{\mathfrak{D}} \mathbf{V}^{\mathrm{In}}(\tau)
  \quad \mathrm{and} \quad  
  \mathcal{T}(\chi:\tau) = \left(\mathbf{V}^{\mathrm{T}}(\chi)\right)^\mathrm{tr} \mathbf{\mathfrak{D}} \mathbf{V}^{\mathrm{In}}(\tau),
  \label{eq13}
\end{equation}
where $\mathbf{V}^{\mathrm{R}}(\chi)$, $\mathbf{V}^{\mathrm{T}}(\chi)$ and $\mathbf{V}^{\mathrm{In}}(\tau)$ are column vectors of length $M(2N+1)$ 
with entries 
\[
  \left[\mathbf{V}^{\mathrm{R}}(\chi)\right]_{(j-1)(2N+1)+N+n+1} = \frac{\cvar^n}{\pi}
  \expo^{\cvar k(x_j\cos\chi - y_j\sin\chi)} \expo^{-\cvar n\chi},
\]
\[
  \left[\mathbf{V}^{\mathrm{T}}(\chi)\right]_{(j-1)(2N+1)+N+n+1} = \frac{(-\cvar)^n}{\pi}
  \expo^{-\cvar k((x_j-L)\cos\chi + y_j\sin\chi)} \expo^{\cvar n\chi},
\]
$1\le j\le M$, $-N\le n\le N$, and
\[
  \left[\mathbf{V}^{\mathrm{In}}(\tau)\right]_{(m-1)(2N+1)+N+s+1} = \cvar^s
  \expo^{\cvar k(x_m\cos\tau + y_m\sin\tau)} \expo^{-\cvar s\tau},
\]
$1\le m\le M$, $-N\le s\le N$, and the superscript tr indicates matrix transpose. 

We obtain a numerical solution by discretising the angular variables $\chi$ and $\tau$ in the range $[-\pi/2,\pi/2]$ with $2N_a+1$ evenly distributed 
samples $\left\{\chi_i = i\pi/2N_a,\,-N_a\le i\le N_a\right\}$. The reflection and transmission kernels then become square matrices 
$\mathbf{\mathfrak{R}}$ and $\mathbf{\mathfrak{T}}$ of size $2N_a+1$ with entries 
\[
\left[\mathbf{\mathfrak{R}}\right]_{p,q} = \mathcal{R}(\chi_{p-N_a-1}:\chi_{q-N_a-1}) \quad \mathrm{and} \quad 
\left[\mathbf{\mathfrak{T}}\right]_{p,q} = \mathcal{T}(\chi_{p-N_a-1}:\chi_{q-N_a-1}),
\]
$1\le p,q\le 2N_a+1$. The discretised reflection and transmission spectra are then calculated from equation \eqref{eq12} using standard numerical 
integration techniques, \eg a trapezoidal rule. 


\section{Results}

The model imposes no assumptions on the FSD in the band, except that the number of floes must be finite. This allows us to define 
the FSD randomly and generate results using ensemble averaging, given a certain ice concentration (relative surface area covered by sea ice). 
However, the computational cost increases significantly with the number of floes, thus limiting the extent of the MIZ that can be considered (solving the 
problem for more than 300 floes generally becomes impractical). On the other hand, the present model is well-suited for the ice band problem, which has a 
limited extent by nature. In this section, we first analyze the effect of both randomness and increasing the number of floes in a simple band setting, and 
will then show results for more realistic ice bands. 

The amplitude spectra $A^{\mathrm{R}}(\chi)$ and $A^{\mathrm{T}}(\chi)$ calculated from equation \eqref{eq12} provide information on both the angular 
distribution and energy intensity of the reflected and transmitted wave components. The total reflected and transmitted energies are given by
\begin{equation}
  E^{\mathrm{R}} = \int_{-\pi/2}^{\pi/2} \left|A^{\mathrm{R}}(\chi)\right|^2 \ud \chi \quad \mathrm{and} \quad 
  E^{\mathrm{T}} = \int_{-\pi/2}^{\pi/2} \left|A^{\mathrm{T}}(\chi)\right|^2 \ud \chi,
  \label{eq14}
\end{equation}
respectively, and are defined relative to the normalized incident energy. Energy conservation in the system yields $E^R+E^T=1$, and provides a useful 
numerical check for convergence with respect to $N_a$. Convergence of the results with respect to the angular sampling 
parameter, $N_a$, and the number of angular modes, $N$, will be assumed throughout, ensuring an accuracy of 4 significant digits. The water depth is 
considered fixed to $h=200\rm\,m$, which provides deep water conditions for wave periods $T\le20\rm\,s$. 

\subsection{Single row: randomness and band extent}

We consider first a single row of identical ice floes (with radius $a$ and thickness $D$) with centres aligned along the $y$-axis and with constant 
centre-to-centre spacing $\mathcal{S}=300\rm\,m$.  We assume that there is an odd number of floes $M=2N_f+1$. The case $N_f=0$ corresponds to a single 
floe centred at the origin and, for positive $N_f$, we add $N_f$ floes on either side of the $x$-axis, so the problem is symmetric with respect to the 
$x$-axis. This regular array defines the reference geometry, from which random arrays are obtained by introducing a perturbation on the radius, thickness 
and position of each floe. We denote by $\epsilon$, a uniformly distributed random parameter taking values between $-1$ and 1. Then for each floe $j$, 
$-N_f\le j\le N_f$, we define the radius $a_j = a + \epsilon \widetilde{a}$, the thickness $D_j = D + \epsilon \widetilde{D}$ and the centre position 
$(x_j,y_j) = (0,j\mathcal{S}) + \epsilon (\widetilde{x}, \widetilde{y})$. The parameters $\widetilde{a}$, $\widetilde{D}$, $\widetilde{x}$ and 
$\widetilde{y}$ control the variance of the associated distributions. In the following, results given for randomized arrays of floes are averaged over 
50 simulations, unless otherwise specified.

\subsubsection{Transmitted energy}

\begin{figure}
  \centering{\includegraphics[width=0.95\textwidth]{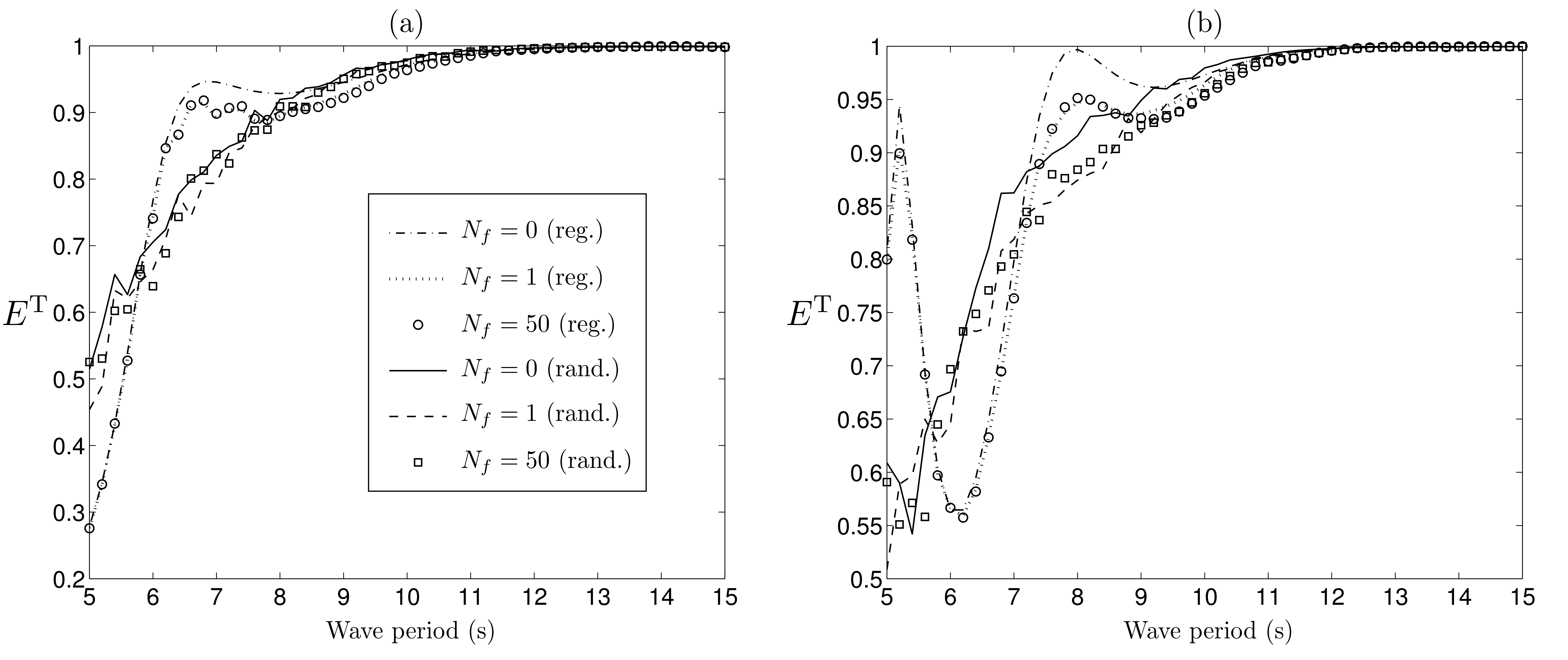}}
  \caption{Transmitted wave energy plotted against wave period for a floe concentration of (a) 20\% and (b) 50\%. We compare the responses of the regular 
  and random arrays, for different values of $N_f$, \ie\ $N_f=0$ (dash-dotted and solid), $N_f=1$ (dotted and dashed) and $N_f=50$ (circles and squares).}
  \label{fig2}
\end{figure}

We analyze the effect of randomizing the FSD and increasing the number of floes on the transmitted energy $E^{\mathrm{T}}$. We consider first an array 
with low concentration ($\approx20\%$). We set $a=75\rm\,m$, $D=1.5\rm\,m$, $\widetilde{a} = 25\,\text{m}$, $\widetilde{D} = 0.5\,\text{m}$ and 
$\widetilde{x} = \widetilde{y} = 50\,\text{m}$. Figure~\ref{fig2}a shows the transmitted wave energy plotted against wave period in the range 
$T=5\text{--}15\rm\,s$, corresponding to wavelengths {40--350\,m}. Results are given for the regular array configuration, \ie\
$\widetilde{a} = \widetilde{D} = \widetilde{x} = \widetilde{y} = 0$, and the random case, with the same mean properties as for the regular 
case. In each case, we vary the number of floes in the band with $N_f=0,\,1$ and $50$. As we could expect, randomizing the FSD and averaging over many 
simulations removes the local maximum observed for the regular case at $T\approx6.8\rm\,s$. The transmitted energy then gradually decreases for shorter 
periods without resonance or near-resonance effects, which characterize the response in the regular array problem. 

Increasing the number of floes in the array, we see that the transmitted energy converges surprisingly fast to the long array response for both regular 
and random cases. The single floe case ($N_f=0$) already provides a reasonable approximation, and for 3 floes ($N_f=1$) the transmitted energy is very 
close to the response for 101 floes ($N_f=50$). This suggests that the spatial repartition of the scattered energy for a single row of floes is mostly 
governed by the single floe response. Multiple interaction effects within the row have little influence on how much wave energy travels through the row. 
However, the situation may be different when we consider multiple adjacent rows (future paper), and will probably require us to take more floes to obtain 
a converged response. 

In figure~\ref{fig2}b, we consider a denser FSD with $a=120\rm\,m$, $D=1.5\rm\,m$, $\widetilde{a} = 10\,\text{m}$, $\widetilde{D} = 0.5\,\text{m}$ and 
$\widetilde{x} = \widetilde{y} = 20\,\text{m}$. In this case, the ice floe concentration is approximately 50\%. The response for the regular array case 
exhibits significant near-resonance features with local maxima at $T\approx5.2$ and $7.9\rm\,s$ and a minimum at $T\approx6.2\rm\,s$. Randomizing the FSD 
removes all these features, as in the previous example. We also observe that the single floe response is the main factor determining the amount of wave 
energy transmitted through the array, and only 3 floes are needed to account for the multiple scattering effects. 

\subsubsection{Directional spectra}

The total reflected and transmitted energy are average quantities over the whole directional spectrum. We now would like to characterize the effect of 
randomizing and increasing the number of floes on the directional spreading properties of the transmitted energy, \ie\ $|A^{\mathrm{T}}(\chi)|^2$. For a 
random configuration, the symmetry with respect to the $x$-axis is broken in general, but taking an ensemble average over many simulations would restore 
the symmetry. To limit the number of simulations required to obtain symmetry, we only randomize the position of the floes, leaving radius and thickness 
fixed, \ie\ $\widetilde{a} = \widetilde{D} = 0$.

\begin{figure}
  \centering{\includegraphics[width=0.95\textwidth]{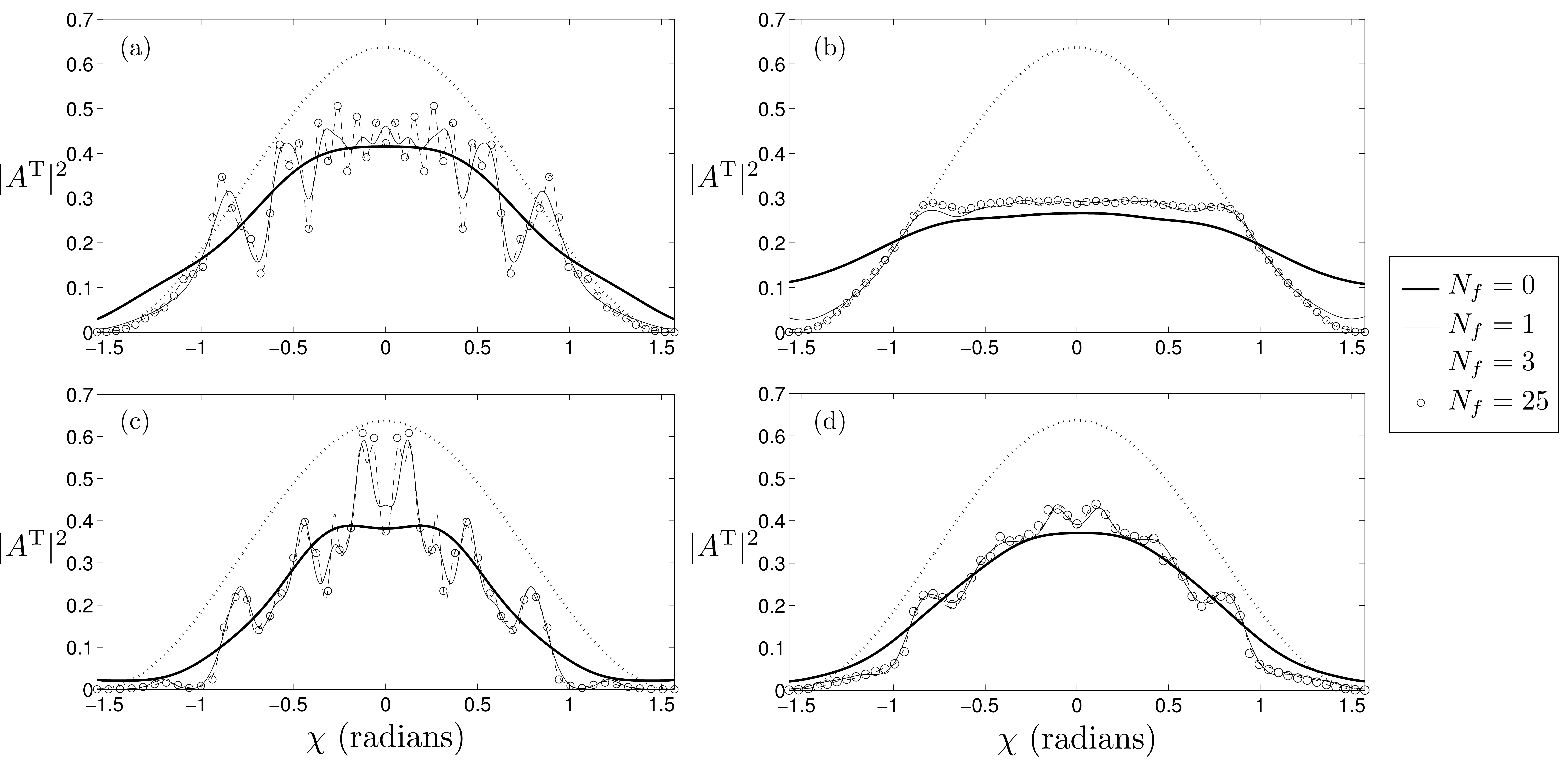}}
  \caption{Directional spectrum of the transmitted energy $|A^{\mathrm{T}}|^2$ for 4 different band extents, $N_f=0,\,1,\,3,\,25$, under an 
  incident spectrum (dotted curve) of period $T=6\rm\,s$. Spectra are given for the regular array configuration in sub-figures (a) and (c), for ice 
  concentrations 20\% and 50\%, respectively. Likewise, the corresponding spectra for random arrays are plotted in sub-figures (b) and (d).}
  \label{fig3}
\end{figure}

We consider the low ice concentration case ($\approx20\%$) first, with $a$, $D$, $\widetilde{x}$ and $\widetilde{y}$, as before. In figures~\ref{fig3}a 
and \ref{fig3}b, the directional spectrum of the transmitted energy $|A^{\mathrm{T}}|^2$ is plotted over the angular range $-\pi/2\le \chi\le \pi/2$, for 
the regular and random array cases, respectively. For each case, we vary the band extent (number of floes in the band), from 1 floe ($N_f=0$) to 51 floes 
($N_f=25$). For the regular array arrangements, increasing the number of floes generates an oscillatory behaviour in the spectrum, which converges to the 
long finite array response quickly, \ie\ for only 7 floes ($N_f=3$). The oscillatory behaviour makes sense, as it is well established that for an infinite 
regular array, the transmitted waves travel only at certain angles, called the scattering angles, determined by the spacing between the floes 
\citep[see, e.g.,][]{peter_etal06}. 

In the random array case, the oscillations disappear due to the averaging process. As a consequence, the convergence is quicker, with only 3 floes 
($N_f=1$) required to approximate the long array response over most of the angular domain. We observe a quasi-flat spectrum in the angular range $(-1,1)$ 
radians, suggesting that an isotropic transmission of the energy occurs there. This means that, statistically, there is no preferred direction of 
propagation in this sub-domain. Outside this angular range, the transmitted spectrum follows the incident spectrum closely, so that the array does not 
influence wave transmission in a statistical sense. 

Figures~\ref{fig3}c and \ref{fig3}d show the same transmitted spectra for an ice concentration of 50\%. For the regular array case, we observe again the 
oscillatory behaviour associated with the scattering angles of the periodic structure, but unlike the low-concentrated example, the oscillations persist 
when we randomise the FSD, although they are are significantly dampened. We conjecture that it is due to the lower variance of the random perturbation 
in the high concentration case ($\widetilde{x}=\widetilde{y}=10\rm\,m$), so that the floes have less freedom to be positioned away from the mean value 
compared to the low concentration arrays, for which we have $\widetilde{x}=\widetilde{y}=50\rm\,m$. The difference in variance also influences the number 
of simulations required to approximate the ensemble average accurately. In practice, the convergence was determined, so the symmetry of the transmitted 
spectra, inherited from the symmetry of the average FSD, would be preserved. For the low concentration case, we had to perform 3000 simulations to obtain 
symmetry, while for the high concentration case, only 500 simulations were required.  

It is remarkable that, again, the long array response can be approximated with only 3 floes for the random array case. However, we expect that more 
floes will be needed for incident directional spectra that are less smooth than the one considered here. In particular, this will be the case when we 
study multiple strips of floes in a subsequent paper.

\subsection{Realistic band}

We now model the propagation of a directional wave spectrum in a band of ice floes, reconstructed from field data provided by \cite{wadhams_etal86}. The 
data are used to approximate a realistic ice band and test it for energy transmission using our model. We will not perform a quantitative comparative 
analysis between experimental data and numerical results in this paper, however, as morphological data are not presented to a sufficient level of detail in the \citeauthor{wadhams_etal86} paper.

The experimental measurements were conducted as part of the MIZEX-84 campaign and took place in the Greenland Sea in June--July 1984. A band experiment 
was performed, in addition to more traditional attenuation experiments in the MIZ. The authors placed a wave buoy on each side of the band to record the 
incoming spectrum and the transmitted spectrum. The ice band extent (in the $y$ direction in our model) was approximately $15\rm\,km$, and its width 
varied between $230\rm\,m$ (narrowest) and $1.3\rm\,km$ (widest). A schematic diagram provided by Wadhams and others suggests that the wave buoys were 
located around the narrowest part of the band, which was composed of about six rows of floes. The FSD is provided after binning the floe properties into 5 
categories for computational purposes. The bins are defined by their characteristic radii $6.25,\,12.5,\,17.5,\,27.5$ and $50\rm\,m$, and proportions of 
the total ice-covered surface area $20,\,30,\,30,\,10$ and $10\%$, respectively. The thickness is set to $D=2\rm\,m$ for all floes. No information is 
given regarding the ice concentration in the band, although we expect that it will be high ($\ge80\%$), as ice edge bands are known to be densely packed 
structures \citep{wadhams83}. 

In our model, we construct the arrangement of floes using a rectangular grid of 6 by 15 square cells. We then randomly place 1 floe per cell from the FSD 
defined earlier, although we discard floes with radii $25.7$ and $50\rm\,m$, as they only account for $1.5\%$ and $0.4\%$ of the total number of floes, 
respectively. Choosing the cell size to be $35\rm\,m$, we obtain an array of floes with 35\% ice concentration. We acknowledge that our rectangular grid 
method is not ideal to define highly concentrated arrangements of floes, but it allows us to include randomness relatively easily. The incident 
directional wave forcing is the same as that considered in the previous section. 

\begin{figure}
  \centering{\includegraphics[width=0.95\textwidth]{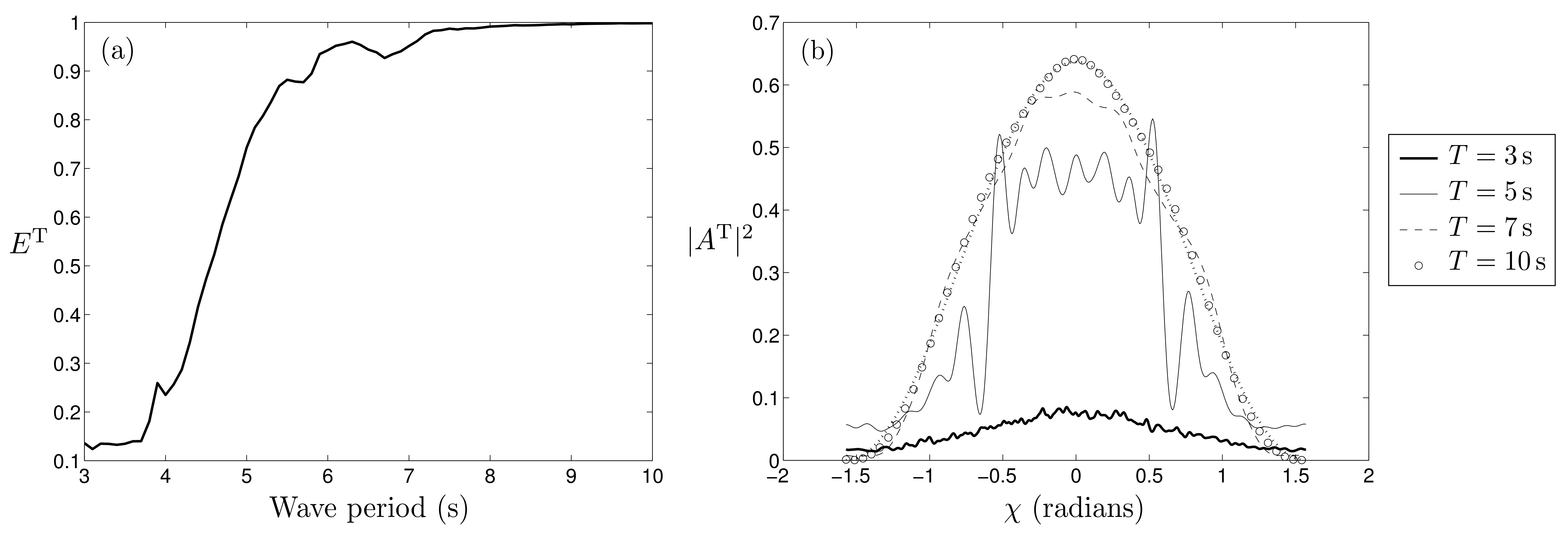}}
  \caption{Analysis of the transmission properties of the wave energy averaged over many random representations the ice band. Sub-figure~(a) 
  shows the total transmitted energy $E^{\mathrm{T}}$ plotted against wave period. In sub-figure~(b), the directional spectrum of the transmitted energy 
  is given for wave periods $T=3$, $5$, $7$ and $10\rm\,s$. The incident spectrum (dotted line) is shown for reference.}
  \label{fig4}
\end{figure}

Figure~\ref{fig4}a shows the total transmitted energy by the ice band in the range of wave periods $T=3\text{--}10\rm\,s$. The plotted curve is the 
averaged transmitted energy from 50 random realizations of the band, specified by the FSD described above. We observe a relatively smooth increase of 
the transmitted energy from about 10\% for short waves to full transmission for long waves. Qualitatively, our results are similar to those provided by 
\cite{wadhams_etal86} --- see their figure~18, as the transition from full reflection to full transmission occurs in the same frequency range. 
Quantitatively, however, our results seem to consistently overestimate the transmitted energy, which is likely to be explained by the difference in ice 
concentration. Higher ice concentration tends to magnify scattering effects and therefore lowers the transmitted wave energy, which is consistent with 
our conjecture for the discrepancy.

We then analyze how the wave directional spectrum is affected after travelling through the ice band. Figure~\ref{fig4}b shows the directional spread of 
the transmitted energy for different values of the wave period, each of which are averaged from 200 simulations. For long waves ($T=7$ and $10\rm\,s$), 
the wave field travels through the band almost unaffected, as suggested by figure~\ref{fig4}a ($E^{\mathrm{T}}>90\%$), and the transmitted spectrum is 
similar to that of the incident waves. For $T=5\rm\,s$, the directional spread of the transmitted energy has an oscillatory nature, similar to what we 
could observe for regular arrays in the previous section. Evidently, the wave field responds to an underlying regular structure of the FSD at this 
frequency, but we cannot provide a conclusive explanation for this behaviour. We note, however, that the wavelength ($\approx39\rm\,m$) is similar to the 
mean spacing between the floes ($35\rm\,m$), which could explain the strong interaction between the wave field and the band. 

For short waves ($T=3\rm\,s$), much of the wave energy has been reflected ($E^{\mathrm{T}}\approx10\%$). We also observe that the transmitted directional 
spectrum becomes almost isotropic, although it still peaks around the normal direction. This phenomenon is typical of wave propagation in the MIZ, as 
reported by \cite{wadhams_etal86}. For such small waves, multiple scattering effects become chaotic within the band, and the ensemble average does not 
favour a particular direction at which the transmitted wave will propagate. For wider regions of sea ice, we expect that all frequency components will 
eventually become isotropic at a certain distance from the ice edge. This effect will be studied in detail in a future paper.

\section{Conclusions}

We devised a new model of wave scattering by a small array of arbitrary circular ice floes, which provides the angular characteristics of the waves reflected 
and transmitted by the array. Important new features are (i) the ability to consider an ambient wave forcing with a realistic continuous directional 
spread and (ii) the possibility to compute the ensemble average of the reflected and transmitted wave properties by randomising the floe size distribution 
in the array. The solution method combines the self-consistent interaction theory for 2D multiple scattering and the plane wave integral representation of 
circular wave components. We then express the reflected and transmitted waves as continuous superposition of plane waves travelling in all directions of 
the angular range. A numerical solution is obtained by discretising the angular spectrum.

We analysed results for a single row of floes perturbed from the regular arrangement. The main findings are summarised as follows:
\begin{enumerate}
 \item computing an ensemble average for the perturbed array over many random simulations eliminates near-resonance features of the transmitted energy and 
 dampens the oscillatory behaviour of its directional spread, both of which are observed in the regular array case. 
 \item The long array response can be approximated surprisingly well with just 3 floes for both regular and random arrays. This suggests that the spatial repartition of the scattered energy is mainly governed by the single floe response, \ie\ multiple scattering has very little effect for the single row 
 geometry. We conjecture that it is a feature of the smooth incident wave spectrum and that more floes will be needed for more perturbed spectra. 
\end{enumerate}

Results were also obtained for a realistic ice band, parametrised using data provided by \cite{wadhams_etal86}. Qualitative agreement has been found for 
the transmitted energy in the swell wave frequency range, while significant discrepancies in ice floe concentration did not allow us to reproduce the 
experimental data quantitatively. For short waves, we showed that the transmitted directional spectrum becomes nearly isotropic, in agreement with 
observations reported by \cite{wadhams_etal86}.

The present paper is a first step towards constructing a cell-based wave/ice interaction model with 2D scattering for use in integrated Arctic system 
models such as TOPAZ \citep{sakov_etal12}. In future studies, our goal will be to extend the single row approach to include multiple strips of ice floes.
This will effectively allow us to study large scale MIZs composed of O(1000) floes, under the same deterministic framework.

\section{Acknowledgements}

The work described in this paper is embedded in the US Office of Naval Research Departmental Research Initiative `Sea State and Boundary Layer Physics of 
the Emerging Arctic Ocean'. The authors are grateful to ONR Award Number N00014-131-0279 and to the University of Otago for financial support. LB 
acknowledges funding support from the Australian Research Council (DE130101571) and the Australian Antarctic Science Grant Program (Project 4123). 


\begin{thebibliography}{21}
\expandafter\ifx\csname natexlab\endcsname\relax\def\natexlab#1{#1}\fi
\expandafter\ifx\csname selectlanguage\endcsname\relax
  \def\selectlanguage#1{\relax}\fi

\bibitem[Bennetts and others, 2010]{bennetts_etal10}
Bennetts, L.~G., M.~A. Peter, V.~A. Squire and M.~H. Meylan, 2010. A three
  dimensional model of wave attenuation in the marginal ice zone, {\em J.
  Geophys. Res.\/}, {\bf 115}, C12043.

\bibitem[Bennetts and Squire, 2008]{bennetts_squire08}
Bennetts, L.~G. and V.~A. Squire, 2008. Wave scattering by an infinite
  straight-line array of axisymmetric floes, {\em Int. J. Offshore Pol.
  Eng.\/}, {\bf 18}, 254--262.

\bibitem[Bennetts and Squire, 2009]{bennetts_squire09}
Bennetts, L.~G. and V.~A. Squire, 2009. Wave scattering by multiple rows of
  circular ice floes, {\em J. Fluid Mech.\/}, {\bf 639}, 213--238.

\bibitem[Bennetts and Squire, 2012]{bennetts_squire12}
Bennetts, L.~G. and V.~A. Squire, 2012. On the calculation of an attenuation
  coefficient for transects of ice-covered ocean, {\em Proc. R. Soc. Lond.
  A\/}, {\bf 468}, 136--162.

\bibitem[Cincotti and others, 1993]{cincotti_etal93}
Cincotti, G., F.~Gori, M.~Santarsiero, F.~Frezza, F.~Furno and G.~Schettini,
  1993. Plane wave expansion of cylindrical functions, {\em Opt. Commun.\/},
  {\bf 95}(4), 192--198.

\bibitem[Frezza and others, 2010]{frezza_etal10}
Frezza, F., L.~Pajewski, C.~Ponti and G.~Schettini, 2010. Scattering by
  dielectric circular cylinders in a dielectric slab, {\em J. Opt. Soc. Am.
  A\/}, {\bf 27}, 687--695.

\bibitem[Jeffries and others, 2013]{jeffries_etal13}
Jeffries, M.~O., J.~E. Overland and D.~K. Perovich, 2013. The Arctic shifts to
  a new normal, {\em Phys. Today\/}, {\bf 66}, 35--40.

\bibitem[Kagemoto and Yue, 1986]{kagemoto_yue86}
Kagemoto, H. and D.~K.~P. Yue, 1986. Interactions among multiple
  three-dimensional bodies in water waves: an exact algebraic method, {\em J.
  Fluid Mech.\/}, {\bf 166}, 189--209.

\bibitem[Martin and others, 1983]{martin_etal83}
Martin, S., P.~Kauffman and C.~Parkinson, 1983. The movement and decay of ice
  edge bands in the winter Bering Sea, {\em J. Geophys. Res.\/}, {\bf 88}(C5),
  2803--2812.

\bibitem[Montiel, 2012]{montiel12}
Montiel, F., 2012. Numerical and experimental analysis of water wave scattering
  by floating elastic plates, (PhD thesis), University of Otago.

\bibitem[Montiel and others, 2013]{montiel_etal13b}
Montiel, F., L.~G. Bennetts, V.~A. Squire, F.~Bonnefoy and P.~Ferrant, 2013.
  Hydroelastic response of floating elastic discs to regular waves. Part 2.
  Modal analysis, {\em J. Fluid Mech.\/}, {\bf 723}, 629--652.

\bibitem[Parkinson and Comiso, 2013]{parkinson_comiso13}
Parkinson, C.~L. and J.~C. Comiso, 2013. On the 2012 record low Arctic sea ice
  cover: Combined impact of preconditioning and an August storm, {\em Geophys.
  Res. Lett.\/}, {\bf 40}, 1356--1361.

\bibitem[Peter and others, 2006]{peter_etal06}
Peter, M.~A., M.~H. Meylan and C.~M. Linton, 2006. Water-wave scattering by a
  periodic array of arbitrary bodies, {\em J. Fluid Mech.\/}, {\bf 548},
  237--256.

\bibitem[Sakov and others, 2012]{sakov_etal12}
Sakov, P., F.~Counillon, L.~Bertino, K.~A. Lisæter, P.~R. Oke and A.~Korablev,
  2012. {TOPAZ4}: an ocean-sea ice data assimilation system for the North
  Atlantic and Arctic, {\em Ocean Sci.\/}, {\bf 8}, 633--656.

\bibitem[Squire and others, 2009]{squire_etal09}
Squire, V.~A., G.~L. Vaughan and L.~G. Bennetts, 2009. Ocean surface wave
  evolvement in the Arctic Basin, {\em Geophys. Res. Lett.\/}, {\bf 36},
  L22502.

\bibitem[Squire and others, 2013]{squire_etal13}
Squire, V.~A., T.~D. Williams and L.~G. Bennetts, 2013. Better operational
  forecasting for the contempary Arctic via ocean wave integration, {\em Int.
  J. Offshore Pol. Eng.\/}, {\bf 23}(2), 81--88.

\bibitem[Wadhams, 1983]{wadhams83}
Wadhams, P., 1983. A mechanism for the formation of ice edge bands, {\em J.
  Geophys. Res.\/}, {\bf 88}(C5), 2813--2818.

\bibitem[Wadhams and others, 1986]{wadhams_etal86}
Wadhams, P., V.~A. Squire, J.~A. Ewing and R.~W. Pascal, 1986. The effect of
  the marginal ice zone on the directional wave spectrum of the ocean, {\em J.
  Phys. Oceanogr.\/}, {\bf 16}, 358--376.

\bibitem[Williams and others, 2013{\natexlab{a}}]{williams_etal13a}
Williams, T.~D., L.~G. Bennetts, V.~A. Squire, D.~Dumont and L.~Bertino,
  2013{\natexlab{a}}. Wave-ice interactions in the marginal ice zone. Part 1:
  Theoretical foundations, {\em Ocean Model.\/}, {\bf 71}, 81--91.

\bibitem[Williams and others, 2013{\natexlab{b}}]{williams_etal13b}
Williams, T.~D., L.~G. Bennetts, V.~A. Squire, D.~Dumont and L.~Bertino,
  2013{\natexlab{b}}. Wave-ice interactions in the marginal ice zone. Part 2:
  Numerical implementation and sensitivity studies along 1D transects of the
  ocean surface, {\em Ocean Model.\/}, {\bf 71}, 92--101.

\bibitem[Young and others, 2011]{young_etal11}
Young, I.~R., S.~Zieger and A.~V. Babanin, 2011. Global trends in wind speed
  and wave height, {\em Science\/}, {\bf 332}(6028), 451--455.

\end{thebibliography}

\end{document}